\documentclass[prl,twocolumn,nofootinbib, preprintnumbers, superscriptaddress]{revtex4-1}

\usepackage{amsmath,amssymb,amscd,simplewick}
\usepackage{listings}
\usepackage{dsfont}
\usepackage{slashed}
\usepackage{color}
\usepackage{ulem}

\usepackage{graphicx}
\usepackage{epstopdf}
\usepackage{subfigure}
\usepackage{epsfig}

\usepackage{xcolor}
\usepackage[colorlinks=true,
            linkcolor=blue,
            urlcolor=blue,
            citecolor=green,          
            bookmarks=true,
            bookmarksnumbered=true,
            breaklinks=true,
            pdfpagemode=Fullscreen,
            pdfstartview=FitBH]{hyperref}

\hypersetup{pdfauthor = {Shao-Feng Ge},
	     pdftitle = {}, 
	     pdfsubject = {}, 
             pdfkeywords = {}, 
	     pdfcreator = {LaTeX with hyperref package},
	     pdfproducer = {dvips + ps2pdf} }

\definecolor{gesfpurple}{rgb}{0.47,0.19,0.42}

\definecolor{gesflanse}{rgb}{0.00,0.50,0.50}

\definecolor{gesfblue}{rgb}{0.08,0.42,0.76}

\definecolor{gesfred}{rgb}{1,0,0}

\definecolor{gesfwhite}{rgb}{1,1,1}

\definecolor{gesfblack}{rgb}{0,0,0}

\newcommand{\geqn}[1]{\hypersetup{linkcolor=blue}(\ref{#1})\hypersetup{linkcolor=blue}}
\newcommand{\gfig}[1]{{\hypersetup{linkcolor=violet}Fig.~\ref{#1}\hypersetup{linkcolor=blue}}}

\newcommand{\nuless}{0 \nu 2 \beta}
\newcommand{\mee}{m_{ee}}

\graphicspath{{figs/}}

\begin{document}


\title{Phenomenological Advantages of the Normal Neutrino Mass Ordering}
\author{Shao-Feng Ge}
\email{gesf@sjtu.edu.cn}
\affiliation{Tsung-Dao Lee Institute, Shanghai Jiao Tong University, China}
\affiliation{School of Physics and Astronomy, Shanghai Jiao Tong University, China}
\author{Jing-yu Zhu}
\email{zhujingyu@sjtu.edu.cn}
\affiliation{School of Physics and Astronomy, Shanghai Jiao Tong University, China}
\affiliation{Tsung-Dao Lee Institute, Shanghai Jiao Tong University, China}

\begin{abstract}
The preference of the normal neutrino mass ordering from the recent cosmological
constraint and the global fit of neutrino oscillation experiments does not seem
like a wise choice at first glance since it obscures the neutrinoless double beta
decay and hence the Majorana nature of neutrinos. Contrary to this naive expectation,
we point out that the actual situation is the opposite. The normal neutrino mass
ordering opens the possibility of excluding the higher solar octant and
simultaneously measuring the two Majorana CP phases in future $\nuless$
experiments. Especially,
the funnel region will completely disappear if the solar mixing angle
takes the higher octant. The combined precision measurement by the JUNO
and Daya Bay experiments can significantly reduce the uncertainty in excluding the
higher octant. With a typical $\mathcal O(\mbox{meV})$ sensitivity on the
effective mass $|\mee|$, the neutrinoless double beta decay experiment can tell
if the funnel region really exists and hence exclude the higher solar octant.
With the sensitivity further improved to sub-meV, the two Majorana
CP phases can be simultaneously determined. Thus, the normal neutrino mass ordering
clearly shows phenomenological advantages over the inverted one.
\end{abstract}

\maketitle 

{\it Introduction} --
The neutrino oscillation \cite{PMNS1,PMNS2} is the first established new physics beyond
the Standard Model (SM) of particle physics \cite{PDG}, although it is not clear whether it is due to a
genuine mass or just an environmental matter effect \cite{scalarNSI,darkNSI,
	talkNSI1,talkNSI2,talkNSI3,Choi:2019zxy}. In the last 20 years, various neutrino experiments
have made impressive progresses by measuring the neutrino mixing angles and
two mass splittings \cite{deSalas:2018bym,Esteban:2018azc}.
The neutrino oscillation (mixing and mass splitting) patterns are
coherently weaved, to be wise after the event. 
In 1995, S. Wojcicki pointed out that there seems to be an intelligent
design of neutrino parameters \cite{Wojcicki},
as a ``{\it light-hearted argument}'' \cite{Goodman}: 1) The solar splitting
$\Delta m^2_s \equiv \Delta m^2_{21} = 7.39^{+0.21}_{-0.20} \times 10^{-5}\,\mbox{eV}^2$ is at the right scale to have 
the MSW resonance \cite{Wolfenstein:1977ue, resonant1, resonant2, resonant3};
2) The solar angle $\theta_s \equiv \theta_{12} = {33.82^\circ}^{+0.78^\circ}_{-0.76^\circ}$ takes the right choice to have
sufficiently large oscillations ($\sim 0.8$) at KamLAND;
3) The atmospheric splitting $\Delta m^2_a \equiv \Delta m^2_{31} = 2.528^{+0.029}_{-0.031} \times 10^{-3}\,\mbox{eV}^2$ allows full
oscillation in the middle range of possible distances travelled by atmospheric neutrinos;
4) The atmospheric angle $\theta_a \equiv \theta_{23} = {48.6^\circ}^{+1.0^\circ}_{-1.4^\circ}$ is big enough so that
oscillations could be easily seen;
5) The reactor angle $\theta_r \equiv \theta_{13} = {8.60^\circ} \pm 0.13^\circ$ is small enough so as not
to confuse the above measurements but nevertheless large enough
to allow the leptonic CP phase and mass ordering (MO) measurements.
The recent T2K and NO$\nu$A data indicates a nearly maximal Dirac CP phase,
$\delta_{\rm D} = {221^\circ}^{+39^\circ}_{-28^\circ}$, which is also a good sign.
All the quoted best-fit and uncertainty values are obtained with the normal
ordering (NO), $m_1^{}<m_2^{}<m_3^{}$.

The only exception comes from the neutrino MO. According to the global
fit \cite{deSalas:2018bym,Esteban:2018azc} and cosmological constraint \cite{RoyChoudhury:2019hls},
NO is preferred \cite{deSalas:2018bym}. This
is especially not understandable, in contrast to the coherent picture of mixing
angles and mass splittings described above. With NO, the neutrinoless double
beta ($\nuless$) decay has a sizeable chance ($\gtrsim 1\%$ for $|\mee| \leq 1\,\mbox{meV}$)
to fall into the funnel region \cite{Ge:2016tfx} and
hence becomes invisible. Even if the
effective mass $|\mee|$ is not inside the funnel region, it is still much more
difficult to measure the $\nuless$ decay with NO.
A naive expectation is that the inverted ordering (IO), $m_3^{}<m_1^{}<m_2^{}$,
is a better choice.
Why make it difficult to measure the Majorana nature of neutrinos after paving
the way for measuring the oscillation patterns? Especially, the Majorana
nature is theoretically well motivated. While the mixing angles and mass splittings are
essentially model parameters \cite{Xing:2019vks},
the Majorana nature is driven by the seesaw
mechanisms \cite{dim51,dim52,seesawI1,seesawI2,seesawI3,seesawI4,seesawI5,seesawI6,seesawII1,seesawII2,seesawII3,seesawII4,Foot:1988aq},
leptogenesis \cite{Fukugita:1986hr}, and charge quantization
\cite{chargeQuantization1,chargeQuantization2}.
If there is an intelligent design behind the established oscillation patterns,
it is hard to imagine that the $\nuless$ decay for measuring the Majorana nature
is left unattended. Thus choosing the NO is hence dubbed
as ``{\it God's Mistake}'' \cite{Goodman}.

This naive expectation is not necessarily true and we provide two
arguments. The NO makes it possible to exclude the higher solar octant
and simultaneously measure the two Majorana CP phases.
Note that the so-called ``{\it intelligent design}" \cite{Wojcicki} and
``God's mistake" \cite{Goodman} are just triggers of our thinking
and
should not be considered as the logic starting point or ingredient of our
scientific argument.
In this paper, we try to explore the phenomenological potentials
of the $\nuless$ decay experiments with the NO, rather than making prediction on
which mass ordering should be correct.

{\it The Solar Octant} --
In the presence of the vector type non-standard interaction (NSI), the
solar octant becomes obscured by the degeneracy with MO, the Dirac CP
phase, and for high energy experiments also the
$\epsilon_{ee}$ element from the vector NSI \cite{Coloma:2016gei}.
To make it clear, we parametrize the neutrino mixing matrix as
$V_\nu = U_{23}(\theta_a) U_{13}(\theta_r) U_{12}(\theta_s, \delta_{\rm D})$ and
the Hamiltonian as
\begin{equation}
  \mathcal H
=
  \frac {V_\nu D^2_\nu V^\dagger_\nu}{2 E_\nu}
+ V_{cc}
\left\lgroup
\begin{matrix}
  1 + \epsilon_{ee} \\
& 0 \\
& & 0
\end{matrix}
\right\rgroup \,,
\label{eq:H}
\end{equation}
where $D_{\nu}^{2} \equiv \mbox{diag}\{- \frac 1 2 \Delta m^2_s, \frac 1 2 \Delta m^2_s, \Delta m^2_a - \frac 1 2 \Delta m^2_s\}$ is the diagonal mass matrix.
Note that parametrizing the Dirac CP phase $\delta_{\rm D}$ in the 1--2 mixing
$U_{12}(\theta_s, \delta_{\rm D})$ is equivalent to the conventional parametrization
\cite{PDG} in the 1--3 mixing, up to a rephasing matrix on each side of $V_\nu$.
For simplicity, only the real $\epsilon_{ee}$ element of the vector NSI is considered
since the others are not relevant.
The vacuum term $\mathcal H_{\rm vac} $ of Eq. \geqn{eq:H}, i.e. the first term on the right side of the equation, changes into $ - \mathcal H^*_{\rm vac}$,
under the transformation: $\sin \theta_s \leftrightarrow \cos \theta_s$,
$\delta_{\rm D} \rightarrow \pi - \delta_{\rm D}$, and
$\Delta m^2_a \rightarrow - \Delta m^2_a + \Delta m^2_s$ \footnote{
The $\pi$ term in the Dirac CP phase transformation contributes the overall
minus sign, together with $\sin \theta_s \leftrightarrow \cos \theta_s$ and
$\Delta m^2_a \rightarrow - \Delta m^2_a + \Delta m^2_s$, while the $- \delta_{\rm D}$
term contributes the complex conjugation. Both terms
are important since non-trivial physical consequences can appear if there is only
one of them \cite{Akhmedov:2001kd}.}.
For the matter potential term, the minus sign comes from
$\epsilon_{ee} \rightarrow - 2 - \epsilon_{ee}$.
Without breaking this degeneracy, the solar mixing angle
has two solutions in the lower or higher octant, respectively.

Although neutrino scattering data can help to break the degeneracy to
some extent \cite{Coloma:2016gei, Coloma:2017ncl, Giunti:2019xpr},
it can only apply to sufficiently heavy mediators. 
The HO (LMA-dark) solution \cite{Miranda:2004nb,Escrihuela:2009up}
	is not uniquely related to heavy mediators but can also be contributed
	by light mediators since their NSI effects are proportional to coupling over mass,
	something like $g^2 / m^2$. By proportionally adjusting coupling and mass,
	there is no particular mass scale for NSI. Especially, for light mediators
\cite{Farzan:2015doa,Farzan:2015hkd,Forero:2016ghr,Babu:2017olk,Farzan:2017xzy,Denton:2018xmq},
it is always possible to tune the coupling $g$ to be small enough to evade
those experimental searches with 
sizeable momentum transfer including the coherent scattering experiments, since
	the propagator $g^2 / (q^2 - m^2) \approx g^2/q^2$ can be highly suppressed
  by the tiny $g$ \cite{Coloma:2017egw,Esteban:2018ppq,Coloma:2019mbs}.

In this paper, we discuss how to exclude the solar HO solution by the $\nuless$
decay measurement with the help of the precision measurement at the reactor neutrino
oscillation experiments, which can apply universally to both light and heavy mediators.
Since the $\nuless$ decay is free of NSI and the reactor neutrino oscillation with
very low energy is not sensitive to matter effects \cite{Li:2016txk} to which the NSI
effect belongs, their
combination can provide an independent check for the aforementioned degeneracy.
In Ref. \cite{N.:2019cot}, the authors have pointed out that the effective mass
has different distributions in the LO and HO cases. Especially the HO case can
be readily measured and sets a new sensitivity goal. Their conclusion also briefly mentioned the possible
`{\it refutal}' of the HO. This section elaborates the aspect of excluding
the HO solution. Especially, we stress the crucial role played by the
precision measurements of reactor neutrino experiments JUNO and Daya Bay
in significantly reducing the uncertainty of relevant oscillation parameters
($\theta_r$, $\theta_s$, $\Delta m^2_a$, and $\Delta m^2_s$).
In addition, we discuss in detail how the cosmological mass sum can also help to exclude 
the HO solution once combined.

The octant transformation, $c_s \leftrightarrow s_s$ where
$(c_x, s_x) \equiv (\cos \theta_x, \sin \theta_x)$, is actually equivalent to
$m_1 \leftrightarrow m_2$. The effective mass $\mee$ for the $\nuless$ decay is,
\begin{equation}
  \mee
=
  c^2_r c^2_s m_1 e^{i \tilde \delta_{\rm M1}}
+ c^2_r s^2_s m_2
+ s^2_r       m_3 e^{i \tilde \delta_{\rm M3}} \,,
\label{eq:mee}
\end{equation}
where $\tilde \delta_{\rm Mi} \equiv \delta_{\rm Mi} - \delta_{\rm D}$ is a combination of
the Majorana CP phase $\delta_{\rm Mi}$ and the Dirac phase $\delta_{\rm D}$. Note that
this form is the same as the conventional parametrization with two complex phases
attached to the $m_1$ and $m_3$ terms. Although the $m_2$ term has no complex phase,
it plays an equal role as the $m_1$ term since both are vectors on the complex plane.
This becomes more transparent by simply rotating the phase $e^{i \tilde \delta_{\rm M3}}$ away
from the $m_3$ term, rendering both the $m_1$ and $m_2$ terms complex. Since the two
Majorana CP phases $\tilde \delta_{\rm Mi}$ are unknown and can take any values,
the effective mass $|\mee|$ distribution is invariant under the combined switch
$c^2_s m_1 \leftrightarrow s^2_s m_2$. The effect of $c_s \leftrightarrow s_s$ is
the same as $m_1 \leftrightarrow m_2$. A direct consequence is that, if
$m_1 \simeq m_2$, the octant transformation $c_s \leftrightarrow s_s$ would leave no
significant consequence in the $\nuless$ decay. Since the two Majorana CP phases
are completely free, the transformation of the
Dirac CP phase, $\delta_{\rm D} \rightarrow \pi - \delta_{\rm D}$, can be easily absorbed into
its Majorana counterparts.
To see the effect of switching the solar octants, the two mass eigenvalues have to
be non-degenerate, which also applies for the
beta decay where the key parameter is
$m_\beta \equiv c^2_r c^2_s m_1 + c^2_r s^2_s m_2 + s^2_r m_3$ \cite{Farzan:2002zq}.

\begin{figure}[t]
	\centering
	\includegraphics[width=0.4\textwidth]{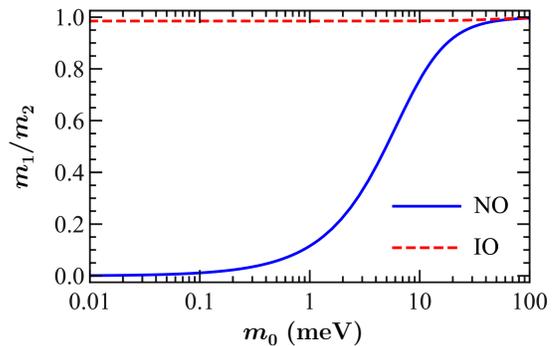}
	\caption{The mass ratio $m_1 / m_2$ for NO and IO.}
	\label{fig:rM12}
\end{figure}

The \gfig{fig:rM12} shows the ratio of $m_1/m_2$ as a function of the lightest mass,
$m_0 \equiv m_1$ for NO and $m_0 \equiv m_3$ for IO.
Since the atmospheric mass splitting is much larger than the solar one,
$\Delta m^2_s / \Delta m^2_a \approx 3\% \ll 1$, $m_1$ and $m_2$ are almost
degenerate across the whole parameter space for IO. In contrast, they can
be non-degenerate for NO. With $m_1 \lesssim 40\,\mbox{meV}$, there is
apparent deviation from being degenerate. The smaller $m_1$, the bigger
the deviation.

\begin{figure}[t!]
	\centering
	\includegraphics[width=0.4\textwidth]{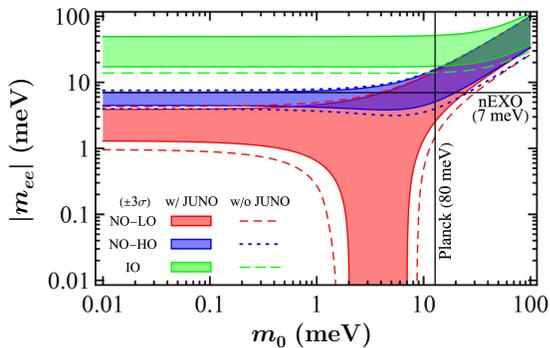}
	\caption{The allowed range of $|m_{ee}|$ for NO with LO (NO-LO, red),
		NO with HO (NO-HO, blue), and IO (green).
    The dashed lines indicate the $3\sigma$ uncertainty according
		to the current global fit \cite{deSalas:2018bym,Esteban:2018azc} of neutrino oscillation
    parameters ($\theta_s$, $\theta_r$, $\Delta m^2_s$, and $\Delta m^2_a$) while
    for the filled region we further impose the projected precision
    of $\sin^2\theta_s$ (0.54\%) and $\Delta m_s^2$ (0.24\%) at the
    future JUNO experiment \cite{An:2015jdp}. For comparison, the typical
		future prospects of the $0 \nu 2 \beta$ decay measurement \cite{Kharusi:2018eqi}
		and cosmological constraint \cite{Dvorkin:2019jgs,RoyChoudhury:2019hls}
		are shown as horizontal and vertical lines, respectively.}
	\label{fig:rangeNO}
\end{figure}

As expected, there is no visible difference between the solar octants for IO
while for NO the effect is sizeable, as shown in \gfig{fig:rangeNO}.
For IO, the predictions with LO and HO almost completely overlap with each other.
So we show only one case in green color and label it  as ``IO''. For
NO, the prediction with LO (in red color and labeled as ``NO-LO'') is totally
different from the one with HO (in blue color and labeled as ``NO-HO'').
Especially, the funnel region for NO-LO completely disappears for NO-HO. Instead,
the effective mass $|\mee|$ is bounded from below across the whole parameter range.
The different effective mass distributions between NO-LO and NO-HO
as well as the degenerate distributions between IO-LO and IO-HO \cite{N.:2019cot}
is actually a reflection of the $m_1$--$m_2$
non-degeneracy or degeneracy, respectively.

\begin{figure}[t]
	\centering
	\includegraphics[width=0.4\textwidth]{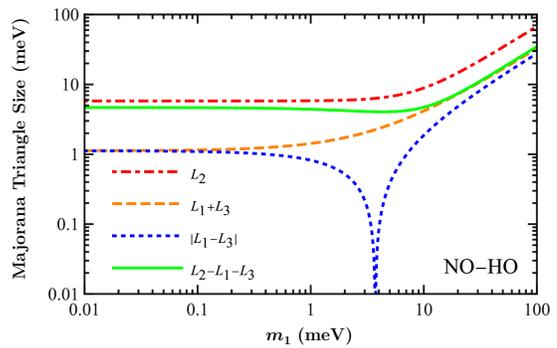}
	\caption{The relevant boundary parameters for NO-HO.}
	\label{fig:boundary}
\end{figure}

According to
the geometrical picture \cite{Xing:2014yka}, the lower and upper limits are completely
determined by the lengths of the three complex vectors,
($L^{\rm LO}_1 \equiv c^2_r c^2_s m_1$,
$L^{\rm LO}_2 \equiv c^2_r s^2_s m_2$, and $L_3 \equiv s^2_r m_3$ for LO).
With $m_1$ and $m_2$ switched, namely $L^{\rm HO}_1 \equiv c^2_r s^2_s m_1$ and
$L^{\rm HO}_2 \equiv c^2_r c^2_s m_2$ for HO, the situation becomes totally
different from the LO case. For convenience, we use only the LO value for
the solar angle, $\theta_s < \pi/4$, globally.
As shown in \gfig{fig:boundary}, $L^{\rm HO}_2 > L^{\rm HO}_1 + L_3$ holds for the whole
parameter space. Consequently, the lower limit of the effective mass is always
$|\mee|^{\rm NO-HO}_{min} = L^{\rm HO}_2 - L^{\rm HO}_1 - L_3$.
Most importantly, $L^{\rm HO}_2$ never crosses with $L^{\rm HO}_1 + L_3$
since interchanging $c_s \approx \sqrt{2/3}$ and $s_s \approx \sqrt{1/3}$ to switch
from LO to HO can significantly amplify $L^{\rm HO}_2$ and suppress
$L^{\rm HO}_1$. This is especially true for small $m_1$
and hence small $m_1/m_2$. Although $L_3$ contains the largest mass eigenvalue
$m_3$, the suppression of $s^2_r$ makes $L_3$ too small to
compensate the difference between $L^{\rm HO}_1$ and $L^{\rm HO}_2$, and hence
the inequality $L^{\rm HO}_2 > L^{\rm HO}_1 + L_3$ always holds.
For comparison, the boundary parameters for NO-LO can be found in Fig.~10b of
\cite{Ge:2016tfx}.

Although the lower boundary for the effective mass $|\mee|$ with NO-HO is established,
the prediction can still receive significant uncertainty from the neutrino oscillation
parameters for both the lower and upper boundaries, shown as the regions between the dashed
curves for the
$3 \sigma$ variations in \gfig{fig:rangeNO}. As argued in similar situations
\cite{Ge:2016tfx, Dueck:2011hu, Ge:2015bfa}, the largest variation comes from the
uncertainties in the solar angle $\theta_s$. This is the place where the intermediate
baseline reactor neutrino experiment JUNO \cite{An:2015jdp} can help.
The precision measurement on the solar angle $\theta_s$ comes from the slow
oscillation modulated by the smaller solar mass splitting
$\Delta m^2_s$ \cite{Ge:2012wj,Ge:2015bfa},
\begin{equation}
  P_{ee}
=
  1
- \cos^4 \theta_r \sin^2 2 \theta_s \sin^2 \Delta_s
+ \cdots \,,
\label{eq:Pee}
\end{equation}
where $\Delta_s \equiv \Delta m^2_s L / 4 E_\nu$ while $\cdots$ stands for the higher
frequency modes modulated by the larger atmospheric mass splitting $\Delta m^2_a$
and its variation $\Delta m^2_a - \Delta m^2_s$. The above Eq. \geqn{eq:Pee} clearly
indicates that the constraint on the solar angle is in the form of
$\sin^2 2 \theta_s = 4 c^2_s s^2_s$, instead of the individual $c_s$ or $s_s$.
The simulations found that $\sin^2 \theta_s$ can be measured with $0.54\%$ precision
\cite{Ge:2012wj,An:2015jdp,Li:2016txk}, from which the uncertainty of the individual $s^2_s$ can
be extracted as
\begin{equation}
  \delta s^2_s
=
  2 c_s s_s \delta \theta_s
=
  \frac {c^2_s s^2_s}{c^2_s - s^2_s}
  \frac {\delta \sin^2 2 \theta_s}{\sin^2 2 \theta_s} \,.
  \label{eq:s12}
\end{equation}
The right-hand side of Eq. \geqn{eq:s12} is invariant under the octant transformation $c_s \leftrightarrow s_s$,
regardless of an overall minus sign. Since the coefficient $2 c_s s_s$ of the
solar angle variation $\delta \theta_s$ is also invariant under the octant
transformation, the absolute uncertainty of the solar angle is not affected,
no matter which octant it rests in. The JUNO experiment precision
on the solar angle is
quite robust against the solar octant degeneracy and we can directly use the
simulated precision from the JUNO Yellow Book \cite{An:2015jdp}.

The filled regions in \gfig{fig:rangeNO} show the $3 \sigma$ range after taking
JUNO into account. Adding JUNO significantly reduces the uncertainty in the predicted
effective mass, which already seems significant in a log scale plot. Especially,
in the vanishing mass limit, $m_1 \rightarrow 0$, the two regions of NO-LO and NO-HO
overlap with each other when taking the current global fit values of the oscillation parameters
and separate from each other after combining the projected JUNO result.
For $m_1 < 0.4~{\rm meV}$, the NO-HO and NO-LO distributions detach from each other.
Since the
lightest mass eigenvalue $m_1$ is negligible in this range, the upper limit for
NO-LO, $|\mee|^{\rm NO-LO}_{\rm max} = L^{\rm LO}_2 + L_3$, and the lower
limit for NO-HO, $|\mee|^{\rm NO-HO}_{\rm min} = L^{\rm HO}_2 - L_3$
are fully determined by the $m_2$ and $m_3$ terms. The difference between these
two limits is, $L^{\rm HO}_2 - L^{\rm LO}_2 - 2 L_3 = c^2_r (c^2_s - s^2_s) m_2 - 2 s^2_r m_3$. Since the ratio of the coefficients
$2 s^2_r /[ c^2_r (c^2_s - s^2_s)] \approx 6 s^2_r \approx 13.4\%$ is
smaller than $m_2 / m_3 \approx \sqrt{\Delta m^2_s / \Delta m^2_a} \approx 17.1\%$,
$|\mee|^{\rm NO-HO}_{\rm min} - |\mee|^{\rm NO-LO}_{\rm max}$ is always positive.
To avoid overlap between the NO-LO and NO-HO regions, the solar angle
cannot be too large,
\begin{equation}
  \cos 2 \theta_s
\gtrsim
  \frac {2 s^2_r \sqrt{\Delta m^2_a}}
				{c^2_r \sqrt{\Delta m^2_s}}
\approx
  26.8\%
\quad \Rightarrow \quad
  \theta_s
\lesssim
  37.2^{\circ} \,,
\end{equation}
where the boundary is more than $4\sigma$ away from the current experimental best-fit
value \cite{deSalas:2018bym,Esteban:2018azc}. In other words, even considering the fact that the best-fit value of $\sin^2 2 \theta_s$ could vary, it is highly unlikely that the NO-LO and NO-HO regions can overlap
in the range of $m_1 \lesssim 0.4~{\rm meV}$. Having
$s^2_s \approx 1/3$ so that the missing solar neutrino measurements consistently
measured $1/3$ of the predicted flux is not just a coincidence.
The solar angle not being too large so that the $\nuless$ decay can
optimize the chance for excluding the solar HO solution
adds one more argument to the advertised intelligent design of neutrino parameters
\cite{Goodman,Wojcicki}.

Since the JUNO experiment can measure
$(\sin^2\theta_s,\Delta m_s^2, \Delta m_a^2)$ with better than $1\%$ precision
\cite{An:2015jdp} and the Daya Bay experiment can measure $\sin^2 2{\theta}_r$ with $3\%$ precision
\cite{Cao:2017drk}, the remaining uncertainty mainly comes from the $\nuless$ decay
measurement itself, including the effective mass sensitivity $\sigma_{|\mee|^2}$ and its
central value $|\mee|^2_{\rm c}$, as well as the uncertainty of the cosmological constraint
on the neutrino mass sum, $\sigma_{\rm sum}$. For both observations,
we assume Gaussian distribution
with central value at zero unless stated otherwise.
The direct observable in $0\nu2\beta$ experiments is the event rate
that follows the exponential law, $N(t)=N_0 e^{-t/T}$, where $T$
is the corresponding lifetime. From the measured signal event number
$\Delta N=N_0^{}\Delta t/T$ within the experimental exposure time
$\Delta t \ll T$, the decay lifetime can be derived, $T = N_0 \Delta t / \Delta N$.
Conventionally, the lifetime can be equivalently denoted as the half-lifetime,
$T_{1/2} \equiv T \ln 2 = 1/(G|M|^2 |m_{ee}|^2)$,
where  $G$ is the phase space factor and $M$ denotes the nuclear matrix element.
The lifetime $T$ is measured experimentally while the phase factor $G$
and the nuclear matrix element come from theoretical calculations.
The effective mass is then obtained as, $|m_{ee}|^2 =1/( G |M|^2 T \ln 2)$.
The major uncertainty comes from the experimental one in the lifetime
measurement and the theoretical one in the nuclear matrix element calculation,
both contributing to the uncertainty $\sigma_{|m_{ee}|^2}$,
\begin{eqnarray}
  P_{\nuless}(|m_{ee}^{}|^2)
=
  \frac 1 {\sqrt{2\pi} \sigma_{|m_{ee}|^2}}
  e^{-\frac{\left( |m_{ee}|^2 - |m_{ee}|^2_{\rm c} \right)^2}{2 \sigma_{|m_{ee}|^2}^2}} 
\end{eqnarray}
For generality, we introduce the central value $|m_{ee}|^2_{\rm c}$. If no
event is observed, the distribution peaks at vanishing $\Delta N$ or
$|m_{ee}|_{\rm c} = 0$. Similarly, we
assume the Gaussian probability distribution of the sum of neutrino masses to be:
\begin{eqnarray}
  P_{\rm cosmo} \left( \sum_i m_i \right)
=
  \frac 1 {\sqrt{2 \pi} \sigma_{\rm sum}}
  e^{-\frac{\left( \sum_i^{} m_i^{} \right)^2}{2 \sigma_{\rm sum}^2}}\;.
\end{eqnarray}
 
\begin{figure}[t]
	\centering
	\includegraphics[width=0.425\textwidth]{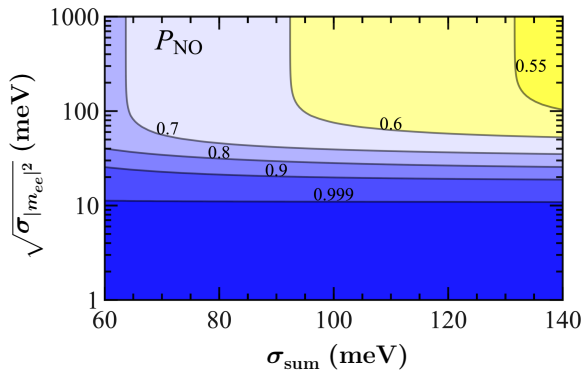}
	\caption{The relative probability of NO as a function of the cosmological sensitivity
		($\sigma_{\rm sum}$) and the $0 \nu 2 \beta$ decay sensitivity ($\sigma_{|m_{ee}|^2}$).}
	\label{fig:probNO}
\end{figure}

As pointed out above, the combined JUNO \cite{An:2015jdp} measurement
and Daya Bay \cite{Adey:2018zwh,Cao:2017drk}
can significantly reduce the uncertainties from the oscillation parameters
to make them negligibly small compared with the uncertainties from the $\nuless$ decay
measurement itself. So we fix the oscillation parameters
($\theta_r$, $\theta_s$, $\Delta m^2_a$, and $\Delta m^2_s$) to their current
best fit values \cite{deSalas:2018bym,Esteban:2018azc} in the following discussions.
The only remaining parameters are just the two
Majorana CP phases ($\tilde \delta_{\rm M1}$ and $\tilde \delta_{\rm M3}$)
and the lightest mass $m_0$. Given a particular mass ordering (NO or IO),
its corresponding likelihood $\mathcal L_{\rm MO}(\sigma_{|m_{ee}|^2}, \sigma_{\rm sum})$ can be evaluated as
\begin{eqnarray}
  \int \hspace{-1mm}
  P_{0\nu2\beta} \left( |m_{ee}|^2_{\rm MO} \right)
  P_{\rm cosmo} \left( \sum_i m_i \hspace{-1mm} \right)
  {\rm d} m_0 \frac{{\rm d}\tilde\delta_{\rm M1}}{2\pi}\frac{{\rm d}\tilde\delta_{\rm M3}}{2\pi} \,,
\quad
\end{eqnarray}
where $m_0 = m_1 (m_3)$ for MO = NO (IO), respectively.
The relative probability
\begin{eqnarray}
  P_{\rm NO,IO}
\equiv
  \frac{\mathcal L_{\rm NO,IO}}{\mathcal L_{\rm NO} + \mathcal L_{\rm IO}}
\end{eqnarray}
quantifies how well the normal (inverted) mass ordering fits the observations,
namely, the NO (IO) sensitivity.
We show  how $P_{\rm NO}$ changes with different $\sigma_{\rm sum}$ and
$\sigma_{|m_{ee}|^2}$ in \gfig{fig:probNO}, assuming no $\nuless$ decay is
observed and hence $|m_{ee}|_c = 0$.
For $\sqrt{\sigma_{|\mee|^2}} \gtrsim 50\,\mbox{meV}$,
the NO sensitivity mainly comes from the cosmological constraint and otherwise
from the $\nuless$ decay. Around
$\sqrt{\sigma_{|\mee|^2}} \sim 50\,\mbox{meV}$, the two mass orderings can
already be distinguished with sensitivity
$P_{\rm NO} \approx 0.7$. In other words, the NO can be identified with
${\cal O}(10\,{\rm meV})$ sensitivity of $\sqrt{\sigma_{|\mee|^2}}$.

\begin{figure}[t]
	\centering
	\includegraphics[width=0.425\textwidth]{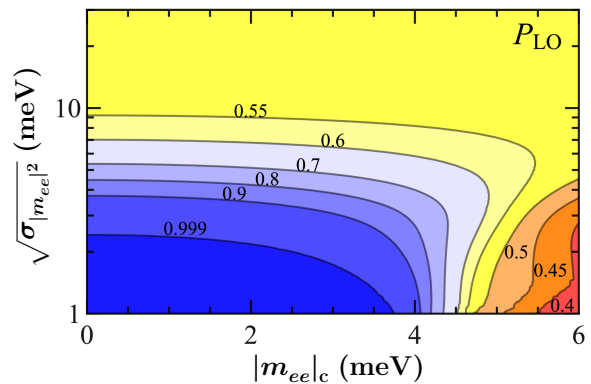}
	\caption{The relative probability of NO-LO from the $0 \nu 2 \beta$ decay effective
		mass sensitivity ($\sigma_{|m_{ee}|^2}$) and its central value ($|m_{ee}|^2_{\rm c}$).}
	\label{fig:probLO}
\end{figure}

After establishing the NO, distinguishing the solar octants takes the similar
definition,
\begin{eqnarray}
  P_{\rm LO,HO}
\equiv
  \frac {{\cal L}_{\rm NO-LO,HO}}
        {{\cal L}_{\rm NO-LO} + {\cal L}_{\rm NO-HO}} \,,
\end{eqnarray}
to quantify the probability that the lower (higher) solar octant is favored.
\gfig{fig:probLO} illustrates the values of $P_{\rm LO}^{}$ with different
$\sqrt{\sigma_{|\mee|^2}}$ and $|m_{ee}^{}|^{}_{\rm c}$.
It is possible to exclude the NO-HO solution if the $\nuless$ decay sensitivity
further improves to $\sqrt{\sigma_{|\mee|^2}} \lesssim 4\,\mbox{meV}$.
According to \gfig{fig:rangeNO}, the lowest
point of the lower boundary for NO-HO is $|m_{ee}| = 3.2\,\mbox{meV}$ at
$m_1 = 5.3\,\mbox{meV}$ without JUNO or $|m_{ee}| = 3.8\,\mbox{meV}$ at
$m_1 = 4.5\,\mbox{meV}$ with JUNO, lower than $4\,\mbox{meV}$ \cite{N.:2019cot,Deepthi:2019ljo}.
However, the realistic measurement has no clear cut. As long as the
$\nuless$ sensitivity $\sqrt{\sigma_{|\mee|^2}}$ goes below $10\,\mbox{meV}$,
which is within the exploration range of future experiments such as
nEXO \cite{Kharusi:2018eqi} and the proposed JUNO-LS detector \cite{Zhao:2016brs},
the possibility for excluding the NO-HO solution can appear: If
the $\nuless$ decay is not observed, the NO-HO solution can be excluded,
with external input of the Majorana nature of neutrinos
\cite{nu-collider,nc-scattering,nu-antinu1,nu-antinu2,nu-antinu3,nu-antinu4,nu-antinu5,nu-antinu6,nu-antinu7,Majorana-EMD1,Majorana-EMD2, Majorana-EMD3,CNB}.
Note that there are already quite a few discussions on the prospect of
the $\mathcal O(\mbox{meV})$ sensitivity of $|m_{ee}|$
\cite{Kharusi:2018eqi,Agostini:2017jim,Xing:2015zha,Ge:2016tfx,Cao:2019hli,Penedo:2018kpc,Revealing}
from both experimental and theoretical perspectives.
 
{\it The Two Majorana CP Phases} -- If the $\nuless$ decay sensitivity further improves
to the sub-meV scale, it is then possible to simultaneously determine the
two Majorana CP phases \cite{Xing:2015zha,Ge:2016tfx,Cao:2019hli}.
The basic logic is that the three complex vectors in Eq. \geqn{eq:mee} form a
closed {\it Majorana triangle} on the complex plane if the effective mass
$|\mee|$ vanishes.
Once the lengths $L_i$ of its three sides are known, its three inner angles can be
uniquely determined as functions of $L_i$. Two of the three inner angles are actually
the two Majorana CP phases as defined in Eq. \geqn{eq:mee}.

Observing the $\nuless$ decay indicates a nonzero effective mass $|\mee|$, corresponding
to only one degree of freedom. Then only one combination of the two Majorana CP
phases can be determined or constrained. But a vanishing
effective mass, $|\mee| = 0$, yields two independent constraints, $\mee = 0$ or
more explicitly, $\mathbb R(\mee) = \mathbb I(\mee) = 0$, where $\mathbb R$ and
$\mathbb I$ extract the real and imaginary components, respectively. Two constraints
can resolve two degrees of freedom, explaining why the
two Majorana CP phases can be simultaneously determined. The same situation can happen
for the more realistic case with some upper limit $U$,
$|\mee| \leq U$, which can convert to two independent upper limits,
$\mathbb R(\mee) \leq U$ and $\mathbb I(\mee) \leq U$. The two Majorana CP phases are
then determined/constrained within some contour.
Again, the JUNO \cite{An:2015jdp} and Daya Bay \cite{Cao:2017drk} experiments
can play an important role by significantly reducing the experimental uncertainties from the
oscillation parameters.

This simultaneous determination of the two Majorana CP phases can only happen when
the effective mass $|\mee|$ falls into the funnel region and hence only for NO.   
With IO, one physical degree of freedom would become invisible forever,
which is a big loss for physics search. In contrast, NO makes it possible to
measure all physical variables without losing any information.
No physical degrees of freedom would be missing.

It seems that the vanishing $|\mee|$ is a disappointing future for the $\nuless$ decay experiments,
which is not necessarily true. The prospect of simultaneously determining the
two Majorana CP phases provides a continuous motivation for improving the experimental
sensitivity. Either we can verify the Majorana nature or measure the two Majorana
CP phases. Both are physically important. To some extent, the $\nuless$
decay has no-loss future.
With other alternative measurements providing the Majorana nature
\cite{nu-collider,nc-scattering,nu-antinu1,nu-antinu2,nu-antinu3,nu-antinu4,nu-antinu5,nu-antinu6,nu-antinu7,Majorana-EMD1,Majorana-EMD2, Majorana-EMD3,CNB}, the $\nuless$ experiment
can simultaneously measure the two Majorana CP phases.

{\it Conclusion} -- We envision the future prospect of neutrino mass ordering and
its role in the $\nuless$ decay by assuming the Majorana nature of  neutrinos. The NO is not the seemingly boring option or
``{\it God's Mistake}'',
but can lead to much more vivid landscapes. First, with $\mathcal O(10\,\mbox{meV})$
sensitivity on the effective mass $|\mee|$, the $\nuless$ decay measurement can
distinguish NO from IO. Second, if the sensitivity further improves to
$\mathcal O(\mbox{meV})$, the $\nuless$ decay measurement can
exclude the solar HO. Different from the NO-LO option that has a funnel
region in the effective mass distribution, the effective mass of the NO-HO option
is bounded from below, $|\mee| \geq 3.2\,(3.8)\,\mbox{meV}$ without (with) input
from JUNO. The solar angle is at the right value to separate
the NO-LO region from the NO-HO one with vanishing or relatively small $m_1$.
Finally, if the sensitivity  improves even further to sub-meV, NO allows the two
Majorana CP phases to be simultaneously determined in the absence of the $\nuless$ decay
signal, observing all physical degrees of freedom.
During this adventure, the input of the solar angle from JUNO and the Majorana nature from independent measurements are necessary. The rich mine in
the $\nuless$ decay is just starting to appear and the global fit preference of NO
is not a nightmare, but an inspiring herald of a new era.

\section*{Acknowledgements}

The work of SFG is supported by JSPS KAKENHI (JP18K13536)  and the Double First Class start-up fund (WF220442604) provided by Shanghai Jiao Tong University.
JYZ is supported by the National Natural Science Foundation of China (11275101
and 11835005). SFG would like to thank the hospitality of KIAS where this paper
was partially finalized. SFG is also grateful to Danny Marfatia for bringing
attention to the NSI degeneracies in the neutrino mass ordering and the solar octant.

\end{document}